\documentclass[nohyperref]{article}

\usepackage[a4paper,
            bindingoffset=0.2in,
            left=1in,
            right=1in,
            top=1in,
            bottom=1in,
            footskip=.25in]{geometry}

\usepackage{microtype}
\usepackage{graphicx}
\usepackage{subfigure}
\usepackage{booktabs}
\usepackage{natbib}
\setcitestyle{authoryear,round,citesep={;},aysep={,},yysep={;}}
\usepackage{parskip}
\usepackage{hyperref}

\usepackage{amsmath}
\usepackage{amssymb}
\usepackage{mathtools}
\usepackage{amsthm}
\usepackage[capitalize,noabbrev]{cleveref}

\theoremstyle{plain}

\theoremstyle{definition}

\theoremstyle{remark}

\author{
   Justin Jude\\
   School of Informatics\\
   University of Edinburgh\\
   Edinburgh, Scotland, EH8 9AB \\
   \texttt{justin.jude@ed.ac.uk} \\
   \and
   Matthew G. Perich \\
   Icahn School of Medicine at Mount Sinai \\
   New York, NY 10029 \\
   \texttt{mperich@gmail.com} \\
   \and
   Lee E. Miller \\
   Feinberg School of Medicine \\
   Northwestern \\
   Chicago, IL 60611\\
   \texttt{lm@northwestern.edu} \\
   \and
   Matthias H. Hennig \\
   School of Informatics\\
   University of Edinburgh\\
   Edinburgh, Scotland, EH8 9AB \\
   \texttt{m.hennig@ed.ac.uk} \\
}
\title{Robust alignment of cross-session recordings of neural population activity by behaviour via unsupervised domain adaptation}
\date{\vspace{-5ex}}
\begin{document}
\maketitle

\begin{abstract}
Neural population activity relating to behaviour is assumed to be inherently low-dimensional despite the observed high dimensionality of data recorded using multi-electrode arrays. Therefore, predicting behaviour from neural population recordings  has been shown to be most effective when using latent variable models. Over time however, the activity of single neurons can drift, and different neurons will be recorded due to movement of implanted neural probes. This means that a decoder trained to predict behaviour on one day performs worse when tested on a different day. On the other hand, evidence suggests that the latent dynamics underlying behaviour may be stable even over months and years. Based on this idea, we introduce a model capable of inferring behaviourally relevant latent dynamics from previously unseen data recorded from the same animal, without any need for decoder recalibration. We show that  unsupervised domain adaptation combined with a sequential variational autoencoder, trained on several sessions, can achieve good generalisation to unseen data and correctly predict behaviour where conventional methods fail. Our results further support the hypothesis that behaviour-related neural dynamics are low-dimensional and stable over time, and will enable more effective and flexible use of brain computer interface technologies.
\end{abstract}

\section{Introduction}

In the brain, stimuli and behaviour can be decoded from the activity of populations of neurons, and it is well established that correlations or co-variations between neurons are a key ingredient in neural population codes \citep{saxena2019towards}. There has been considerable success developing methods for decoding external variables from recordings of even modestly sized populations of 10s or 100s of neurons \citep{Pandarinath2017InferringAuto-encoders,hurwitz2021building}, raising hopes that brain computer interfaces (BCIs) can be an effective assistive technology for severely disabled patients. However, a decoder, once trained, requires stable recordings to perform well. Over the course of days and weeks, the signals recorded from implanted extracellular probes will inevitably change and drift due to factors such as impedance changes, gliosis and probe and brain movement \citep{chestek2011long}. Non-invasive systems such as electromyography (EMG) sensors will not be worn permanently and positioned slightly different every time, creating even stronger variations in recorded signals. Moreover, the activity of individual neurons can change considerably over similar time scales due to neural plasticity \citep{rule2019causes}. Together these fluctuations will lead to degradation of decoder performance over time, thus to be effective, frequent recalibration of BCI systems would be inevitable.

Given the limited long-term stability of recorded neural signals, reports of relatively stable behaviour decoding over days with the same decoder may seem surprising \citep{chestek2007single}. Recent work by  \citep{Gallego2020Long-termBehavior} however showed that some aspects of the population activity of cortical neurons remain very stable even over months and years. Specifically, this study showed that neural population activity in the primary motor cortex is highly restricted to and evolves along a low-dimensional manifold that is stable even when single neuron activity constantly fluctuates. 

Low-dimensional neural dynamics can be effectively extracted from neural population activity with latent variable models \citep{hurwitz2021building}. These models use an often small number of latent variables (or factors) together with an appropriate observation model that relates latent variables to the recorded activity. Importantly, the latent variables in such models often predict stimuli or behaviour very well even when they were only optimised to reproduce neural activity \citep{hurwitz2021building}. Nonlinear state space models such as LFADS are particularly powerful in predicting single trial activity and behaviour in test data \citep{Pandarinath2017InferringAuto-encoders}.

Therefore, instabilities in neural recordings can be successfully compensated for by re-training the part of a model that translates neural activity into the latent dynamics, which are assumed stable over time \citep{Degenhart2020StabilizationActivity}. As a behaviour decoder, this can be more data-efficient than re-training a decoder from scratch, but still requires regular interventions. Here we ask if it is possible to recover the latent dynamics without any re-training. 

Our approach uses a domain adaptation inspired solution. We treat each recording session as a separate domain, each of which can be used to predict the same set of behaviours. The model is then optimised using both recorded activity and behaviour to recover the same latent variables irrespective of the domain, and is capable of predicting behaviour correctly for a previously unseen session without need for re-calibration. In contrast, latent variable models without domain adaptation fail to generalise to unseen data, and instead partition the latent space into distinct parts corresponding to the individual recording sessions. We test this model with long-term recordings from the primate motor cortex during a reach task and show that, provided sufficient training data, it can predict behaviour well for previously unseen sessions. BCI decoders that can generalise well to unseen sessions or subjects without any re-training have not yet been demonstrated. We believe this is the first work to show such cross-session decoder generalisation without recalibration.

\section{Related Work}
This issue of neural stability is investigated by \citet{Gallego2020Long-termBehavior} where the dynamics of a set of a single animal’s M1 cortex neurons are recorded from over many days. The authors find that the underlying dynamics of these neurons over time are indeed reconcilable. Principal component analysis (PCA) is used to reduce the dimensionality of the neural activity on each day, and these variables are then aligned using canonical-correlation analysis (CCA). After alignment, neural activity is regenerated for up to 16 days with close similarity and accurate decoding of behaviour.

\citet{Farshchian2019AdversarialInterfaces} take this approach a step further and utilise an adversarial approach with a non-linear model (ADAN) to directly align neural activity over many days in order to accurately predict EMG during movement. A discriminator network is trained in a similar fashion to LFADS, tasked with autoencoding neural activity from day 0. A generator neural network is optimised to align neural population activity to that recorded at day 0. The autoencoding discriminator is tasked with maximising the alignment loss. Both of the above approaches successfully align latent variables of many days of seen neural activity which are ultimately used to stably predict behaviour from seen sessions. 
 
\citet{Sussillo2016MakingVariability} build a robust decoder capable of utilising large amounts of training data and maintaining decoding performance in the face of recording condition changes such as neuron turnover. All of the previously mentioned models require neural data from all recording sessions for behaviour decoding to be possible and do not generalise to unseen sessions. \citet{Wen2021RapidModelling} uses adversarial generative modelling to generate large amounts of synthetic spike data from just the behaviour of a separate recording session or subject, mimicking the spike data of that session/subject. Together with this generated synthetic spike data and a small amount of real spike data from the unseen session, the authors are able to achieve relatively good behaviour decoding accuracy on the held out session. This model is more data efficient than the three previously mentioned methods. Nevertheless, some neural spike data from an unseen session is still required for good behaviour decoding.
 
\citet{hurwitz2021targeted} combines ideas from \citet{Pandarinath2017InferringAuto-encoders} and \citet{Sani2021ModelingIdentification} to jointly model the neural activity and external behavioural variables by separating the latent space into behaviourally
relevant and behaviourally irrelevant components; the relevant dynamics are used
to reconstruct the behaviour through a flexible linear decoder and both sets of
dynamics are used to reconstruct the neural activity through a linear decoder with
no time lag. This work shows that an LFADS-like model can jointly model neural activity and associated behaviour or movement.
 
Domain adaptation broadly aims to predict classes from labelled data of a similar nature, albeit from differing sources or domains. The method relevant to this work is by \citet{Ganin2015UnsupervisedBackpropagation}, who use a negative gradient between a domain classifier and feature extractor in order to coerce the feature extractor to produce domain invariant features from which a label predictor can infer data classes reliably. This method of domain unification is unsupervised.

\citet{gonschorek2021removing} use domain adaptation to align data across experiments of two-photon imaging recordings using an autoencoder model and a domain classifier. The authors successfully align their recording sessions but they do not test efficacy on unseen sessions. They also explicitly use experimental session ID as domains and show efficacy on non-sequential data in this respect. In this work we show that it is beneficial to not explicitly use session/experiment ID for domain adaptation but to instead use neural patterns directly to align recording sessions for high dimensional sequential data. 
 
 In this work we model each session of neural recording as a separate domain and predict behaviour from all of these sessions simultaneously. Domain-invariant latent variables are obtained using the paradigm of unsupervised domain adaptation via a negative gradient, which are then optimised to reconstruct the observed behaviour.

\section{Model}
\begin{figure*}[ht!]
\begin{center}
\includegraphics[width=0.8\textwidth]{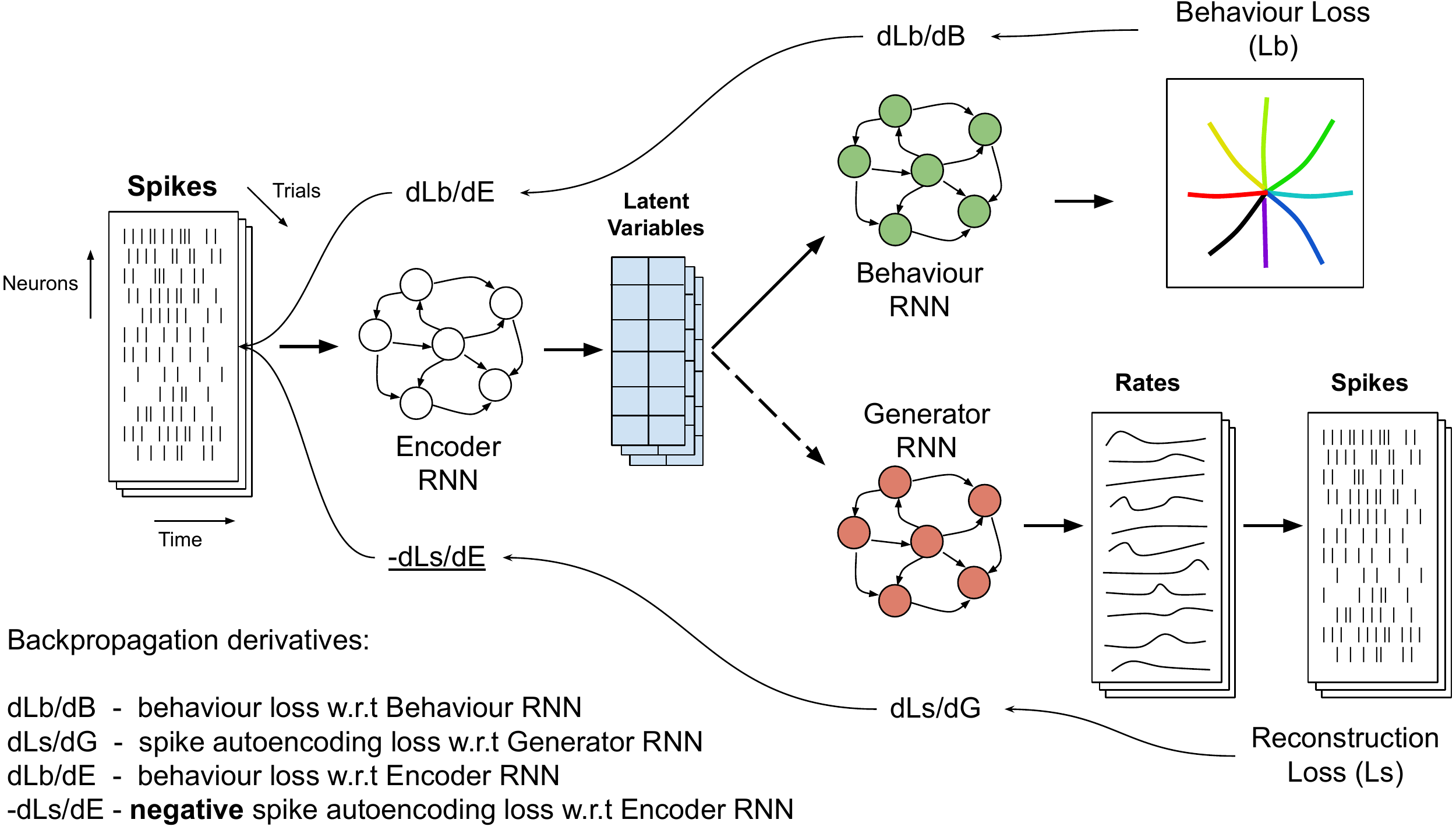}
\caption[lfadsDA]{Our model (SABLE) consists of a sequential variational autoencoding approach combined with a sequential behaviour decoder. Notably, we implement a reverse gradient layer between the neural decoder and encoder GRUs. The encoder can then learn to extract the invariant latent dynamics determining behaviour from data obtained in different sessions with variability in the recorded neural activity. }
\label{fig:lfadsDAmodel}
\end{center}
\end{figure*}

This model is based on the hypothesis that behaviour $y$ is encoded in a stable latent space with variables $z$, and that the two are related linearly as $y=f(z$). Equally, neural activity $x$ is related to the latent variables through a simple function, and as in related models we choose a linear read-out with a Poisson link function to generate non-negative firing rates \citep{Pandarinath2017InferringAuto-encoders}. However, this function will differ between recording sessions (or domains) $d$ as we expect to observe different neurons in each session, and the activity of neurons may change over time. The problem is thus to find the correct encoding function $z=g(x)$ to transform neural activity into the latent space which then allows decoding of behaviour. As explained above, re-training this part of the model for each session can successfully align different sessions. Here we show that this can be achieved without the need for re-training.

Specifically, as proposed by \cite{Pandarinath2017InferringAuto-encoders} we assume that the latent dynamics evolve autonomously provided a set of initial conditions $z_i$ that are modelled as Gaussian random variables. These latent variables are produced for each trial by an encoder network consisting of bidirectional Gated Recurrent Units \citep{Cho2014LearningTranslation} (GRU). They are used to simultaneously predict behaviour, and to reconstruct the original trial-specific neural activity.  We apply recurrent and kernel regularisation to the encoder GRU to enable better generalisation to unseen sessions.

A further bidirectional GRU is used as a decoder for neural reconstruction and a final separate GRU is used to predict behaviour from the generated latent variables. Training is based on a mean squared error loss for behaviour and Poisson likelihood for neural activity. Importantly, we reverse the backpropagation gradient between the neural reconstruction decoder and the encoder. This gradient reversal layer leads to maximisation of the neural reconstruction loss in the encoder network while, at the same time, the neural decoder network is adversarially optimised to minimise neural reconstruction loss. This implicitly encourages the encoder to generate latent variables which are not separated by session of data collection. The behaviour decoder meanwhile forces the encoder to generate latent variables which are differentiated by behaviour. Ultimately, this produces a latent space separable by behaviour but not by session of data collection. The complete model is illustrated in Figure \ref{fig:lfadsDAmodel}.

The model is trained using real neural activity which corresponds to consistent behaviours (movement directions in a centre-out reach task, see below). 
The generative process of our model is as follows:
\begin{gather} \label{eq:gen}
z_i = W_{enc}(\text{GRU}_{\theta_{enc}}(x_{i,1:T})),\\g_{1:T} =  \text{GRU}_{\theta_{dec}}(z_i), \\ b_{1:T} = \text{GRU}_{\theta_{beh}}(z_i),\\  \nonumber \\
r_t =  exp(W_{rate}(W_{fac}(g_t))),\\
\bar{x}_t \sim \text{Poisson}(r_t),\\ \bar{y}_t = W_{beh}(b_t)
\end{gather}

where $\theta_{enc}$, $\theta_{dec}$, $\theta_{beh}$ are the parameters of the GRUs used to encode spike trains into latent variables, decode spike trains from the generated latent variables, and to decode behaviour from the latent variables respectively. $W_{enc}$, $W_{fac}$, $W_{rate}$ and $W_{beh}$ are non-linear layers which produce latent variables, neural activity factors, generate firing rates and predict behaviour respectively at each time step per trial. 

At each training iteration the following three losses are optimised using Adam \citep{Kingma2015Adam:Optimization} asynchronously:

\begin{gather} \label{eq:recon_loss}
L_{rec} = \sum_{t=1}^{t} \log(\text{Poisson}(x_{i,t}\vert r_t))\\
\label{eq:beh_loss}
L_{beh} = \frac{1}{T}\sum_{t=1}^{t}(y_{i,t} - \bar{y}_{i,t} )^2\\
 \nonumber L_{kl} = D_{KL}[\text{GRU}_{\theta_{enc}}(z_i \vert x_i) \vert \vert \mathcal{N}(0, I)]\\ 
 = - \frac{1}{2}[\log(z_{i,\sigma}^2) - z_{i,\mu}^2 - z_{i,\sigma}^2 + 1]
\end{gather}

where $y_i$ is the true behaviour per trial and $\bar{y}_i$ is the predicted behaviour. The loss in eq. \ref{eq:recon_loss} is maximised by the encoder network and minimised by the neural decoder network (and not applicable to the behaviour decoder network). This adversarial training is the most crucial aspect of our model. As the encoder maximises the neural reconstruction loss throughout training, it produces increasingly spike pattern-invariant latent variables. Behaviour loss (eq. \ref{eq:beh_loss}) is minimised by both the encoder and behaviour decoding network while the Kullback–Leibler (KL) divergence loss (between a multivariate standard Gaussian distribution and the encoder generated latent variables) is minimised by just the encoder network. Thus the total error for all parameters in the model across all training trials can be summarised as:

\begin{equation} \label{eq:all_error}
\begin{aligned}
& E(\theta_{enc},W_{enc},\theta_{dec},W_{fac},W_{rate}, \theta_{beh},W_{beh}) = \\ & \sum_{i=1..N}^{i} L_{beh}^i(\theta_{enc},W_{enc},\theta_{beh},W_{beh}) \\ &
+ L_{rec}^i(\theta_{dec},W_{fac},W_{rate}) \\ + \ & \lambda_{kl} L_{kl}^i(\theta_{enc},W_{enc}) - \lambda_r\sum_{i=1..N}^{i}L_{rec}^i(\theta_{enc},W_{enc}) 
\end{aligned}
\end{equation}

where $\lambda_{kl}$ is the weight of KL divergence and $\lambda_r$ is the weight of the reverse gradient applied to the encoder RNN. $\lambda_{kl}$ rises exponentially as training progresses while $\lambda_r$ decays exponentially (thereby increasing session invariance over training). We denote our model Stable Alignment of Behaviour through spike-invariant Latent Encoding (SABLE).

\section{Data}

\subsection{M1 neural recordings during reach task}

\begin{figure}[h!]
\begin{center}
\includegraphics[width=0.45\textwidth]{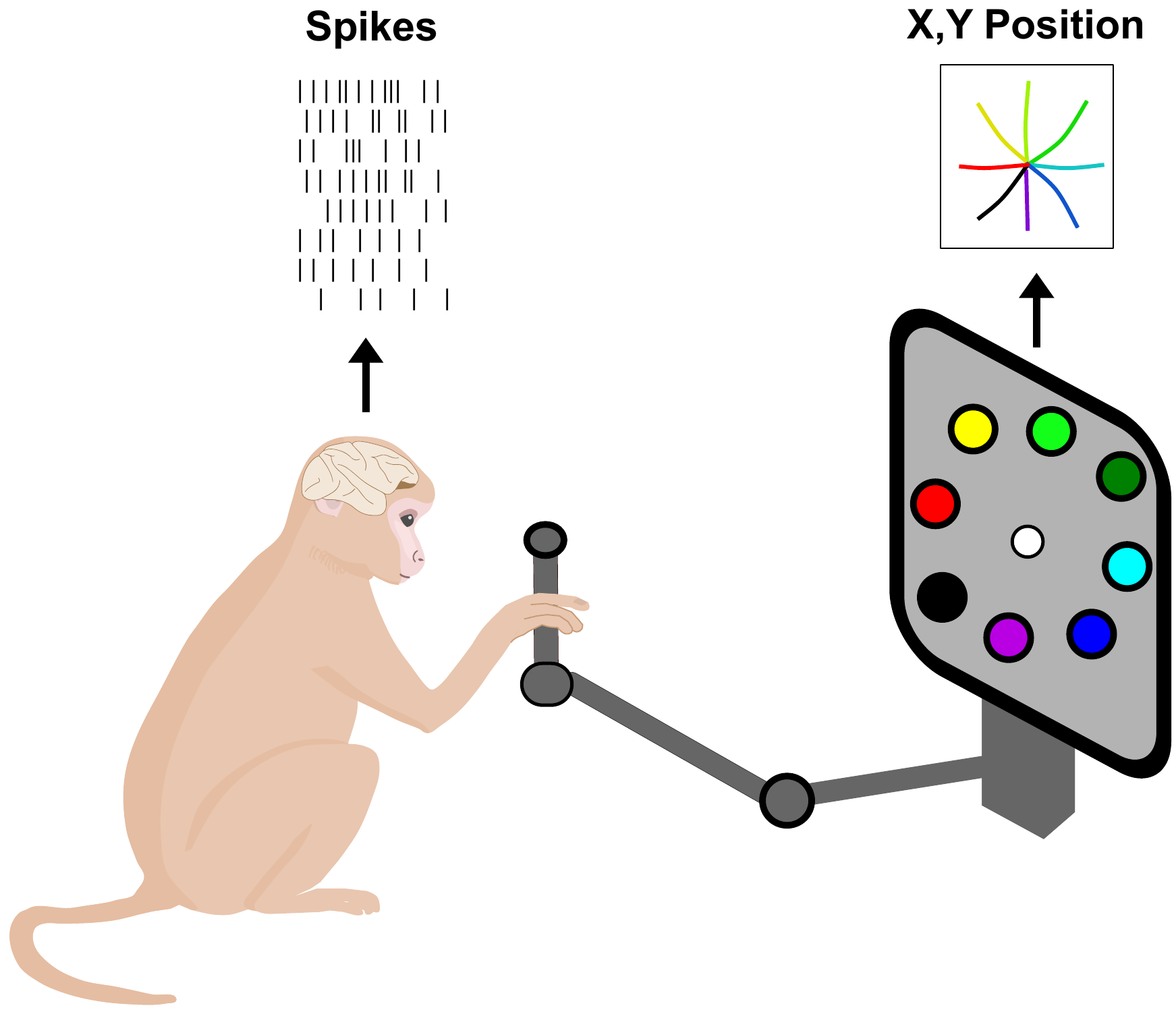}
\caption[experimentsetup]{Experimental setup:  In each trial one randomly chosen target direction (indicated by one of 8 coloured circles) appears on screen, and the monkey is instructed to control the cursor (white circle) by moving the manipulandum. The monkey moves the cursor to the target location after a go cue. The collected data for each trial consists of the neural spikes and monkey hand position across all timesteps. Our model is tasked with predicting hand position from neural spikes at each timestep.}
\label{fig:experiment}
\end{center}
\end{figure}

We verify that SABLE is able to predict behaviour from unseen neural activity by applying it to data from a previously published experiment \citep{Gallego2020Long-termBehavior}. In this experiment, two monkeys are trained to perform a center-out reach task towards eight outer targets. On a go cue, each monkey moves a manipulandum along a 2D plane to guide a cursor on a screen to the target location (Figure \ref{fig:experiment}). On successful trials a liquid reward is given. Spiking activity from the motor cortex (M1) along with the 2D hand position are recorded during each trial. Spike trains are converted into spike counts in 10ms bins, and behaviour variables are used at the same resolution. Only successful trials are used, all trials are aligned to movement onset and cut from movement onset to the shortest reach time across all trials.

For our analysis, we train SABLE on many consecutive days of recorded data and test on a subsequent held out day of recordings for each monkey.
In total there are 13 near consecutive days of recordings for monkey 1 and 6 near-consecutive days of recordings for monkey 2, with fewer recorded neurons and timesteps for monkey 2 overall. Each day for each monkey consists of one recording session.

\section{Models for comparison}
We compare the ability of SABLE to predict behaviour from sessions of unseen spike data against existing methods and against a variation of our own model. We look at 
the following existing models: LFADS \citep{Pandarinath2017InferringAuto-encoders} and RAVE+ \citep{gonschorek2021removing}. We also compare against our own model where we do not reverse the gradient between the encoder and decoder, which we denote SABLE-noREV. In addition, we compare against a baseline RNN (GRU) with a linear readout layer optimised to predict movement from spiking data without  autoencoding. 

LFADS has been shown to have good efficacy at neural reconstruction across trials and sessions with some separation of behaviour in its latent space in previous work. We implement RAVE+ as an autoencoding model with GRUs for the encoder and decoder as our data are time series, and treat recording sessions as separate domains. As with our own model, the encoder is tasked with generating a small number of latent variables following a multivariate standard Gaussian distribution from neural data while the decoder is tasked with reconstructing the data from the latent variables. We use a non-linear layer as a domain classifier on the latent space between the encoder and decoder and implement a negative gradient between this classifier and the encoder network, thus encouraging the encoder to produce session-invariant latent variables. For all models we use the same regularisation techniques in the encoder or predictor as we do for SABLE to maximise generalisation.

For LFADS and RAVE+ we use a separately trained GRU to predict behaviour from the latent space of these models. We do not include ADAN \citep{Farshchian2019AdversarialInterfaces} or the generative adversarial model by \citet{Wen2021RapidModelling} as both require at least some training data from held out session or subject to be effective. Implementation details of SABLE and all comparison models can be found in the Appendix (Section \ref{app:modeldetails}).

\section{Results}

\subsection{Application to motor cortex neural recordings during a reach task}
\begin{figure*}[h!]
\begin{center}
\includegraphics[width=0.85\textwidth]{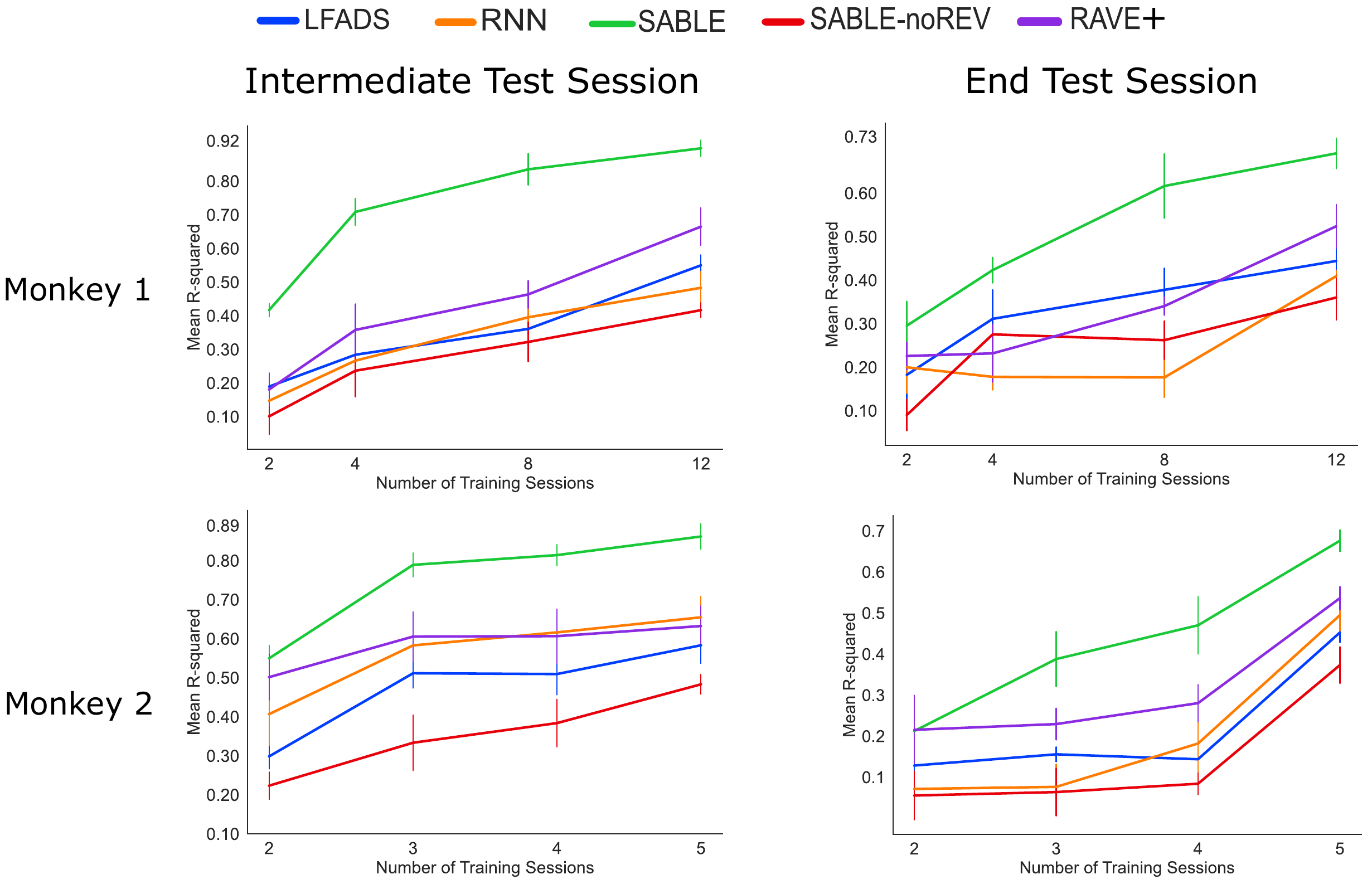}
\caption[Results]{Behaviour prediction performance when testing all models on a completely unseen recording session. We report the mean r-squared between the inferred and true x,y positions. Each model is tested on a held out session while trained on different numbers of training sessions for both monkeys. The left column shows results for a held-out testing session which is in the chronological centre of the training sessions whereas the right column shows results for a test session recorded after all training sessions. Each test condition is run 10 times with different random seeds, with error bars showing standard deviation.}
\label{fig:results}
\end{center}
\end{figure*}

We train all models on varying numbers of training sessions and for both monkeys, testing behaviour (2D hand position) prediction on intermediate and subsequent held out recording sessions. Our results, summarised in Figure \ref{fig:results}, show that SABLE is capable of generalising to unseen data provided a sufficient number of training sessions are provided. In all cases tested SABLE outperforms the comparison models. For example, decoding accuracy for SABLE on an unseen intermediate session for monkey 1 with 12 training sessions is 0.91, which exceeds all other models by at least 0.25. For comparison, the RNN decoder typically yields an accuracy of around 0.92 on held-out data when trained and tested on a single session, indicating that SABLE can achieve saturation performance on unseen data.  RAVE+ has a relatively high decoding performance when a large number of training sessions are used, likely because its domain adaptation paradigm removes some session specific variance in this case. In contrast, SABLE-noREV and LFADS have low decoding performance across monkeys and session numbers although they gradually improve with increasing session numbers.

Comparing the performance between the two monkeys, we see generally better overall decoding performance for monkey 2 at the same number of train sessions as monkey 1, although monkey 1 has far more total training data available (12 total consecutive sessions from monkey 1 vs. 5 for monkey 2) and so has higher peak behaviour decoding performance for all models. In addition, we limit the number of neurons for each monkey to the lowest number of neurons in any given session. Therefore, monkey 1 has 42 neurons of neural data across sessions whereas monkey 2 has 16.

Next we compare the difference between test performance for all models on both monkeys for different  held out test session ordering (intermediate or end). While SABLE achieves end test session decoding performance exceeding that of current methods (0.71 mean r-squared with 12 train sessions), performance on any given intermediate test session is substantially higher (0.90 mean r-squared with 12 train sessions). Moreover, the performance of SABLE decreases noticeably faster when applied to an end test session when the number of training sessions is reduced versus an intermediate test session. This confirms that drift in recordings is gradual, not random.

\subsection{Latent space analysis}
\begin{figure*}[h!]
\begin{center}
\includegraphics[width=1.0\textwidth]{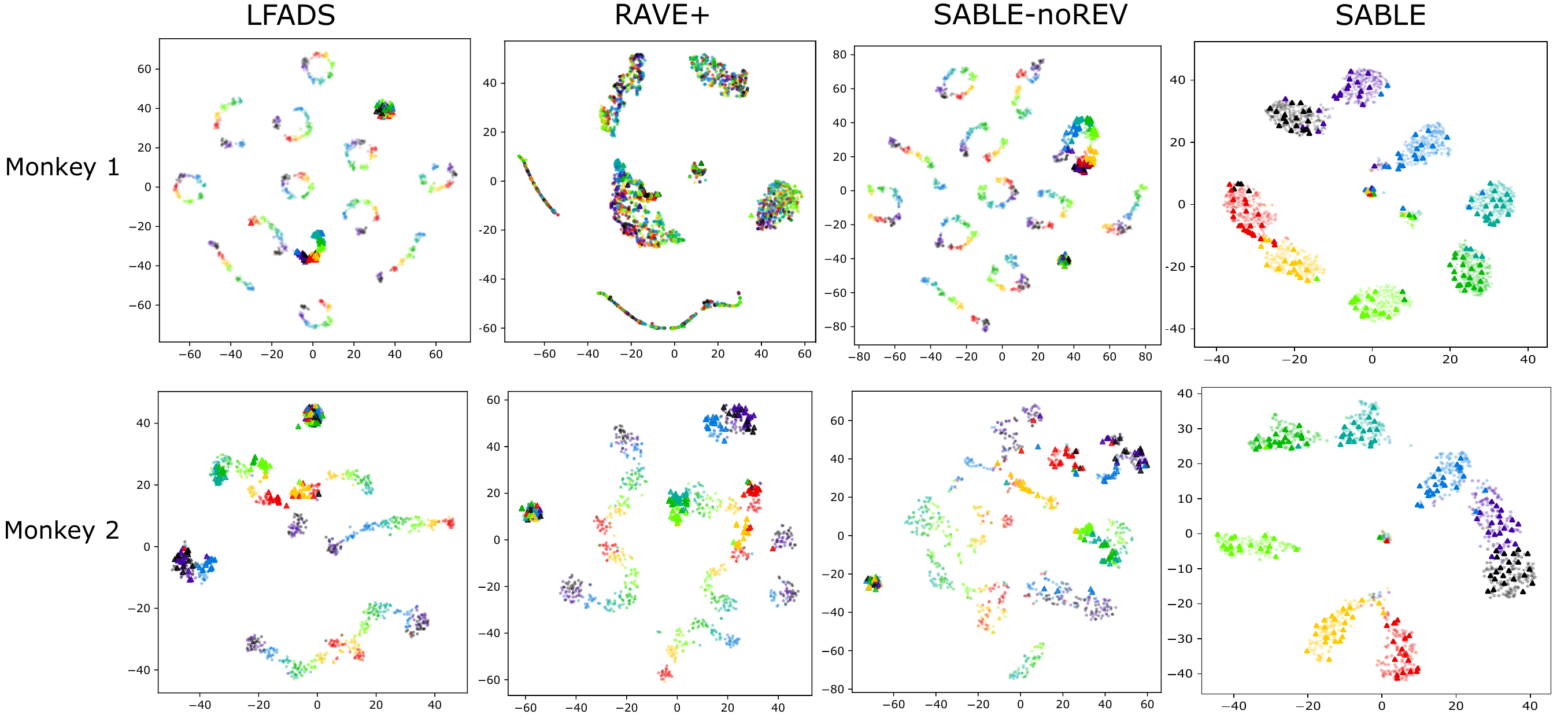}
\caption[tsneplots]{T-SNE embeddings of latent space for autoencoder models. In each embedding, points denoted by a circle are trials from 12 training sessions for monkey 1 and 5 training sessions for monkey 2. Points denoted by a triangle are trials from a held out intermediate test session for both monkeys. Each colour represents a target direction for the centre out reach task.}
\label{fig:tsneplots}
\end{center}
\end{figure*}
Figure \ref{fig:tsneplots} shows T-SNE embeddings of the latent space of all autoencoding models. Each colour represents a different target direction with respect to behaviour, and embeddings of the training data are shown as circles while the test trials are shown as triangles. 

The embeddings of the LFADS latent space show a clear separation that corresponds to the different training sessions. This shows that there are indeed significant differences between the sessions that are captured in the latent space and prevent generalisation to unseen sessions. Within each session cluster there is a good separation by target direction, indicating that the latent variables extract behaviourally relevant information from the neural activity. In contrast, many trials from the test session form a cluster in a region not covered by the training data, with some degree of separation by behaviour in monkey 1 (where more sessions are available for training). Here the model fails to assign these trials to meaningful latent variables as the differences in activity are too large to be mapped appropriately. We denote trials such as these hereafter as unassigned. Some trials are assigned locations in the latent space also occupied by the training data, indicating that despite session differences occasionally a matching of unseen to training data can be achieved, again with some separation by target direction. However there is also a cluster of unseparated test trials which the encoder of the model has failed to produce meaningful latent variables for due to these test trials being too disparate from the training trials. 

For LFADS applied to monkey 2, the encoder still manages to separate trials by train session, but to a lesser degree. We suspect this is due to fewer neurons being available for the sessions of monkey 2. Here there is no longer a cluster of separated test trials, instead some test trials are assigned to existing train clusters as they are fairly similar. Here again however we see a large cluster of unassigned test trials. Overall we see that LFADS clusters train trials and coinciding test trials well but its encoder cannot effectively generate latent variables for dissimilar test trials for behaviour decoding.

The picture is different for RAVE+, where the latent space seems better aligned (more session invariant) but no longer well separated by behaviour for monkey 1 (larger training set and more neurons). In contrast, for monkey 2 we see little session alignment but better behaviour separation is achieved (fewer sessions and fewer recorded neurons). In this case there is some degree of merging of session clusters and organisation by behaviour target direction, but this is insufficient for good test behaviour decoding. For monkey 1 we see 7 clusters, 4 of which have some separation by direction for train session trials. However test trials are not separated well at all by direction. For monkey 2 there is far less clustering by session and some of these clusters separate by direction. Here there is also a large cluster of unassigned test trials. The domain adaptation method used in RAVE+ (reverse gradient based on session ID explicitly) thus does not seem to prevent trials clustering by session.

SABLE-noREV, our model without the reverse gradient, produces a result very similar to LFADS. For both monkeys, there is a cluster of well separated test trials and also many test trials that are unassigned to any cluster (either by direction or session). Therefore, using the latent space for both neural reconstruction and behaviour decoding simultaneously, as proposed by \cite{hurwitz2021targeted}, is not beneficial to test behaviour decoding across sessions.

Finally, applying SABLE to either monkey produces latent spaces which are very well separated by behaviour and almost entirely training-session invariant. When applied to monkey 1 we see a small degree of misclassification of test trials by direction (9\% of total test trials), but only when the correct and incorrect target directions are adjacent to each other spatially in the task outlined in Figure \ref{fig:experiment}. This confirms that more similar behaviours in a task have more similar neural patterns and may be difficult for any decoder to disentangle. There are also a small number of unclassified test trials (3\% of total test trials) in the centre of the embedding plot, we suspect these may be trials with highly contrasting spiking patterns to any train trial. When applied to monkey 2, we see less misclassification (4\% of total test trials) by test behaviour direction and just a couple of unclassified test trials. We also note that both SABLE embeddings are topographically similar and correspond to the spatial aspect of the movement directions of the task outlined in Figure \ref{fig:experiment}.

The stark differences in latent variables seen between SABLE-noREV and SABLE are quite surprising considering the only difference between these models is the reverse gradient between neural decoder and encoder in SABLE versus a positive gradient in SABLE-noREV. This shows the importance of a negative neural reconstruction gradient in training the SABLE encoder network to generate session invariant latent variables. In addition, we suspect that SABLE's encoder generates far fewer unassigned latent variables than the other autoencoding models due to the simpler and more behaviourally structured latent space.

\subsection{Behaviour decoding}

\begin{figure}[h!]
\begin{center}
\includegraphics[width=0.45\textwidth]{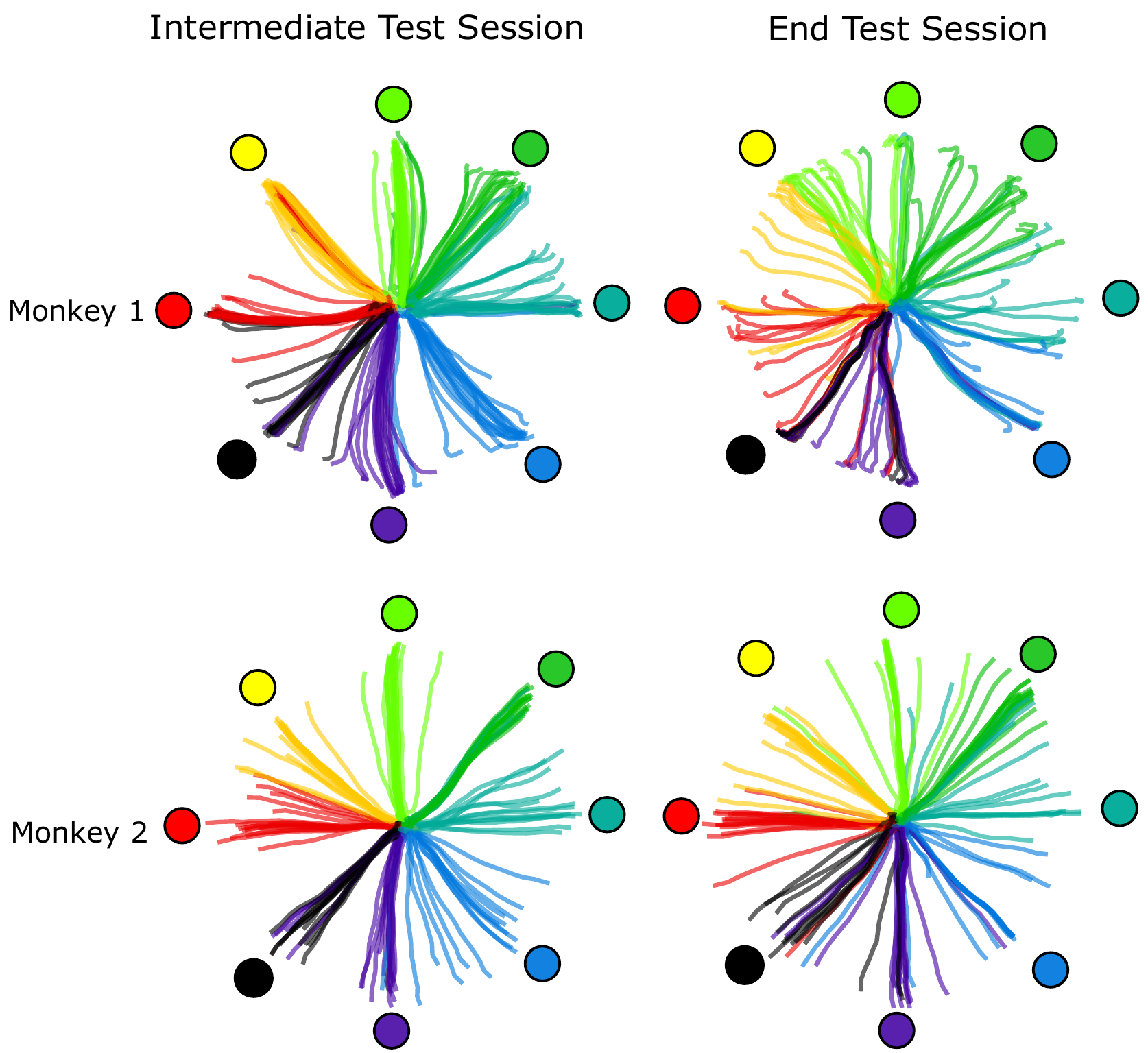}
\caption[behaviour]{Predicted 2D monkey hand position of test trials using SABLE when trained on 12 train sessions for monkey 1 and 5 train sessions for monkey 2 and tested on an unseen intermediate or end test session.}
\label{fig:behaviour}
\end{center}
\end{figure}

Examining the decoded behaviour of monkey hand position using SABLE (Figure \ref{fig:behaviour}) shows good overall reconstruction of movement trajectories, especially when testing on an intermediate test session. The intermediate test session behaviour decoding for both monkeys mirrors the SABLE T-SNE embedding in Figure \ref{fig:tsneplots}. Test trials which are incorrectly assigned with respect to  movement directions (Figure \ref{fig:tsneplots}) are decoded correspondingly (Figure \ref{fig:behaviour}). Therefore, the behaviour decoder network of SABLE directly utilises latent variables in a particular cluster and decodes one particular direction of movement. Our model is thus consistent with the hypothesis outlined above.

When decoding from an end test session however, this phenomenon is less pronounced as the encoder seems to be less certain of the clusters formed in the latent space. There are more wrongly assigned test trials and the decoded movement trajectories are more spread out, leading to a lower overall mean r-squared when predicting behaviour.

\subsection{Predicting behaviour from an unseen subject}
Next we predict behaviour from the unseen neural data from 37 sessions of monkey 2 when SABLE is trained on 14 recording sessions of monkey 1. We use all available recording sessions available for monkey 1 spread across 3 years to train SABLE as we believe this gives the best opportunity for cross subject generalisation. However, for the held-out data we only obtain a mean r-squared of 0.03, so the model fails to generalise to a different animal. 
Examining the T-SNE embedding in Figure \ref{fig:crossmonkeytsne} shows that the trials from the training sessions of monkey 1 cluster well by movement direction but the trials from monkey 2 do not map to these direction clusters as the separated sessions of either just monkey 1 or monkey 2 do (as seen in Figure \ref{fig:tsneplots}). Therefore it appears that the relationship between the recovered latent dynamics and behaviours differs between the two animals, and may require an extra alignment step.

\section{Discussion}

In this work we present a new method, SABLE, for aligning neural activity with complex temporal dynamics from different recording sessions to allow for consistent behaviour decoding across sessions. We apply it to neural recordings from primate motor cortex during a reaching task where the considerable variability between recording sessions prevents generalisation for a conventional decoder.

The model is trained as a variational autoencoder similar to LFADS \citep{Pandarinath2017InferringAuto-encoders}, with an additional gradient from behaviour decoder that disentangles the latent space to enable improved behaviour decoding \citep{hurwitz2021targeted}. Reversing the gradient from the neural reconstruction encourages the model to ignore variability in the activity that is irrelevant for decoding activity, which in turn results in an session-invariant encoding of behaviour-relevant factors. 

Unlike other domain adaptation methods, our model does not require domain labels, but is trained on a single data set that contains different experimental sessions. This is an advantage in potential BCI applications as variability may not only exist between single sessions, but also within a single session, and in addition the degree of variability may differ as well. As a result, the model still requires considerable amounts of session data for good behaviour reconstruction. We found that performance did not saturate when it was trained on 12 sessions. In contrast, good behaviour decoding with our baseline RNN model could be achieved from a single session with less than 200 trials. Yet we expect that the number of trials per session required is much less than used here.

A main limitation of this method, which may also limit its direct application in a BCI system, is that it assumes that behaviour is generated by autonomous neural dynamics which relies on specification of appropriate initial conditions that form the latent variables in the model. This approach has been shown to successfully capture neural dynamics in a range of scenarios \citep{Pandarinath2017InferringAuto-encoders} and has the advantage of a relatively compact and behaviorally relevant latent encoding that supports discovering invariant features in the neural activity. A possible extension to remove this limitation may be the inclusion of a controller input that models additional temporal dynamics to better account for behavioural variability \citep{Pandarinath2017InferringAuto-encoders}. This extension of the latent space can be trained in the same manner and may allow modelling of more complex and variable behavioural paradigms.


We compare our model to RAVE+ \citet{gonschorek2021removing}, to our knowledge currently the only other method for domain adaptation of inter-session data. RAVE+ does show some indication of alignment when sufficient individual recording sessions are available, but its latent space fails to capture behaviourally relevant structure. As a result, behaviour decoding for unseen test data is poor. We suspect that the RAVE+ fails because the temporal dynamics in our data are too variable between trials. As pointed out by the authors, RAVE+ requires consistent temporal dynamics between trials, which can be controlled in experiments where stimuli are presented, but that are rarely obtained in behavioural experiments. The other models shown here (LFADS, RNN decoder) are included to contrast domain adaptation to conventional encoders, and not as a baseline for generalisation performance.

Our results are consistent with recent reports showing that motor control is based on low-dimensional latent neural dynamics that are very stable over time despite ongoing neural drift \citep{Gallego2020Long-termBehavior}. Our model can be  used to discover these latent dynamics in data with high variability. Tests we performed on synthetic data (a Lorenz system) indicate that this approach is also successful when neural dynamics are generated from latent dynamics with random transformations (not shown). 

Our finding that SABLE has a better performance for intermediate held-out sessions than for sessions at the end of a sequence of training sessions suggests that performance will likely eventually decline for long time intervals between train and test sessions. As long as the latent dynamics are stable however, we expect that training the model with more sessions will eventually stabilise generalisation performance.  Taken together these results are encouraging for BCI application as they suggest highly consistent recordings may not be required for good performance as long as it is possible to recover relevant latent dynamics.



\bibliography{biblio,references}

\begin{thebibliography}{17}
\providecommand{\natexlab}[1]{#1}
\providecommand{\url}[1]{\texttt{#1}}
\expandafter\ifx\csname urlstyle\endcsname\relax
  \providecommand{\doi}[1]{doi: #1}\else
  \providecommand{\doi}{doi: \begingroup \urlstyle{rm}\Url}\fi

\bibitem[Chestek et~al.(2007)Chestek, Batista, Santhanam, Byron, Afshar,
  Cunningham, Gilja, Ryu, Churchland, and Shenoy]{chestek2007single}
C.~A. Chestek, A.~P. Batista, G.~Santhanam, M.~Y. Byron, A.~Afshar, J.~P.
  Cunningham, V.~Gilja, S.~I. Ryu, M.~M. Churchland, and K.~V. Shenoy.
\newblock Single-neuron stability during repeated reaching in macaque premotor
  cortex.
\newblock \emph{Journal of Neuroscience}, 27\penalty0 (40):\penalty0
  10742--10750, 2007.

\bibitem[Chestek et~al.(2011)Chestek, Gilja, Nuyujukian, Foster, Fan, Kaufman,
  Churchland, Rivera-Alvidrez, Cunningham, Ryu, et~al.]{chestek2011long}
C.~A. Chestek, V.~Gilja, P.~Nuyujukian, J.~D. Foster, J.~M. Fan, M.~T. Kaufman,
  M.~M. Churchland, Z.~Rivera-Alvidrez, J.~P. Cunningham, S.~I. Ryu, et~al.
\newblock Long-term stability of neural prosthetic control signals from silicon
  cortical arrays in rhesus macaque motor cortex.
\newblock \emph{Journal of Neural Engineering}, 8\penalty0 (4):\penalty0
  045005, 2011.

\bibitem[Cho et~al.(2014)Cho, Van~Merri{\"{e}}nboer, Gulcehre, Bahdanau,
  Bougares, Schwenk, and Bengio]{Cho2014LearningTranslation}
K.~Cho, B.~Van~Merri{\"{e}}nboer, C.~Gulcehre, D.~Bahdanau, F.~Bougares,
  H.~Schwenk, and Y.~Bengio.
\newblock {Learning phrase representations using RNN encoder-decoder for
  statistical machine translation}.
\newblock In \emph{EMNLP 2014 - 2014 Conference on Empirical Methods in Natural
  Language Processing, Proceedings of the Conference}, 2014.
\newblock ISBN 9781937284961.
\newblock \doi{10.3115/v1/d14-1179}.

\bibitem[Degenhart et~al.(2020)Degenhart, Bishop, Oby, Tyler-Kabara, Chase,
  Batista, and Yu]{Degenhart2020StabilizationActivity}
A.~D. Degenhart, W.~E. Bishop, E.~R. Oby, E.~C. Tyler-Kabara, S.~M. Chase,
  A.~P. Batista, and B.~M. Yu.
\newblock {Stabilization of a brain–computer interface via the alignment of
  low-dimensional spaces of neural activity}.
\newblock \emph{Nature Biomedical Engineering}, 4\penalty0 (7), 2020.
\newblock ISSN 2157846X.
\newblock \doi{10.1038/s41551-020-0542-9}.

\bibitem[Farshchian et~al.(2019)Farshchian, Gallego, Miller, Solla, Cohen, and
  Bengio]{Farshchian2019AdversarialInterfaces}
A.~Farshchian, J.~A. Gallego, L.~E. Miller, S.~A. Solla, J.~P. Cohen, and
  Y.~Bengio.
\newblock {Adversarial domain adaptation for stable brain-machine interfaces}.
\newblock In \emph{7th International Conference on Learning Representations,
  ICLR 2019}, 2019.

\bibitem[Gallego et~al.(2020)Gallego, Perich, Chowdhury, Solla, and
  Miller]{Gallego2020Long-termBehavior}
J.~A. Gallego, M.~G. Perich, R.~H. Chowdhury, S.~A. Solla, and L.~E. Miller.
\newblock {Long-term stability of cortical population dynamics underlying
  consistent behavior}.
\newblock \emph{Nature Neuroscience}, 23\penalty0 (2), 2020.
\newblock ISSN 15461726.
\newblock \doi{10.1038/s41593-019-0555-4}.

\bibitem[Ganin and Lempitsky(2015)]{Ganin2015UnsupervisedBackpropagation}
Y.~Ganin and V.~Lempitsky.
\newblock {Unsupervised domain adaptation by backpropagation}.
\newblock In \emph{32nd International Conference on Machine Learning, ICML
  2015}, volume~2, 2015.

\bibitem[Gonschorek et~al.(2021)Gonschorek, H{\"o}fling, Szatko, Franke,
  Schubert, Dunn, Berens, Klindt, and Euler]{gonschorek2021removing}
D.~Gonschorek, L.~H{\"o}fling, K.~P. Szatko, K.~Franke, T.~Schubert, B.~A.
  Dunn, P.~Berens, D.~A. Klindt, and T.~Euler.
\newblock Removing inter-experimental variability from functional data in
  systems neuroscience.
\newblock In \emph{Thirty-Fifth Conference on Neural Information Processing
  Systems}, 2021.
\newblock URL \url{https://openreview.net/forum?id=lVmIjQiJJSr}.

\bibitem[Hurwitz et~al.(2021{\natexlab{a}})Hurwitz, Kudryashova, Onken, and
  Hennig]{hurwitz2021building}
C.~Hurwitz, N.~Kudryashova, A.~Onken, and M.~H. Hennig.
\newblock Building population models for large-scale neural recordings:
  Opportunities and pitfalls.
\newblock \emph{Current Opinion in Neurobiology}, 70:\penalty0 64--73,
  2021{\natexlab{a}}.

\bibitem[Hurwitz et~al.(2021{\natexlab{b}})Hurwitz, Srivastava, Xu, Jude,
  Perich, Miller, and Hennig]{hurwitz2021targeted}
C.~Hurwitz, A.~Srivastava, K.~Xu, J.~Jude, M.~Perich, L.~Miller, and M.~Hennig.
\newblock Targeted neural dynamical modeling.
\newblock \emph{Advances in Neural Information Processing Systems}, 34,
  2021{\natexlab{b}}.

\bibitem[Kingma and Ba(2015)]{Kingma2015Adam:Optimization}
D.~P. Kingma and J.~L. Ba.
\newblock {Adam: A method for stochastic optimization}.
\newblock In \emph{3rd International Conference on Learning Representations,
  ICLR 2015 - Conference Track Proceedings}, 2015.

\bibitem[Pandarinath et~al.(2017)Pandarinath, O’Shea, Collins, Jozefowicz,
  Stavisky, Kao, Trautmann, Kaufman, Ryu, Hochberg, Henderson, Shenoy, and
  Sussillo]{Pandarinath2017InferringAuto-encoders}
C.~Pandarinath, D.~O’Shea, J.~Collins, R.~Jozefowicz, S.~Stavisky, J.~Kao,
  E.~Trautmann, M.~Kaufman, S.~Ryu, L.~Hochberg, J.~Henderson, K.~Shenoy, and
  D.~Sussillo.
\newblock {Inferring single-trial neural population dynamics using sequential
  auto-encoders}.
\newblock \emph{Inferring single-trial neural population dynamics using
  sequential auto-encoders}, 2017.
\newblock ISSN 1548-7091.
\newblock \doi{10.1101/152884}.

\bibitem[Rule et~al.(2019)Rule, O’Leary, and Harvey]{rule2019causes}
M.~E. Rule, T.~O’Leary, and C.~D. Harvey.
\newblock Causes and consequences of representational drift.
\newblock \emph{Current Opinion in Neurobiology}, 58:\penalty0 141--147, 2019.

\bibitem[Sani et~al.(2021)Sani, Abbaspourazad, Wong, Pesaran, and
  Shanechi]{Sani2021ModelingIdentification}
O.~G. Sani, H.~Abbaspourazad, Y.~T. Wong, B.~Pesaran, and M.~M. Shanechi.
\newblock {Modeling behaviorally relevant neural dynamics enabled by
  preferential subspace identification}.
\newblock \emph{Nature Neuroscience}, 24\penalty0 (1), 2021.
\newblock ISSN 15461726.
\newblock \doi{10.1038/s41593-020-00733-0}.

\bibitem[Saxena and Cunningham(2019)]{saxena2019towards}
S.~Saxena and J.~P. Cunningham.
\newblock Towards the neural population doctrine.
\newblock \emph{Current Opinion in Neurobiology}, 55:\penalty0 103--111, 2019.

\bibitem[Sussillo et~al.(2016)Sussillo, Stavisky, Kao, Ryu, and
  Shenoy]{Sussillo2016MakingVariability}
D.~Sussillo, S.~D. Stavisky, J.~C. Kao, S.~I. Ryu, and K.~V. Shenoy.
\newblock {Making brain-machine interfaces robust to future neural
  variability}.
\newblock \emph{Nature Communications}, 7, 2016.
\newblock ISSN 20411723.
\newblock \doi{10.1038/ncomms13749}.

\bibitem[Wen et~al.(2021)Wen, Yin, Furlanello, Perich, Miller, and
  Itti]{Wen2021RapidModelling}
S.~Wen, A.~Yin, T.~Furlanello, M.~G. Perich, L.~E. Miller, and L.~Itti.
\newblock {Rapid adaptation of brain–computer interfaces to new neuronal
  ensembles or participants via generative modelling}.
\newblock \emph{Nature Biomedical Engineering}, 2021.
\newblock ISSN 2157846X.
\newblock \doi{10.1038/s41551-021-00811-z}.

\end{thebibliography}
\bibliographystyle{abbrvnat}

\newpage
\appendix
\onecolumn
\section{Cross-subject decoding T-SNE Embedding}

\begin{figure}[h!]
\begin{center}
\includegraphics[width=0.75\textwidth]{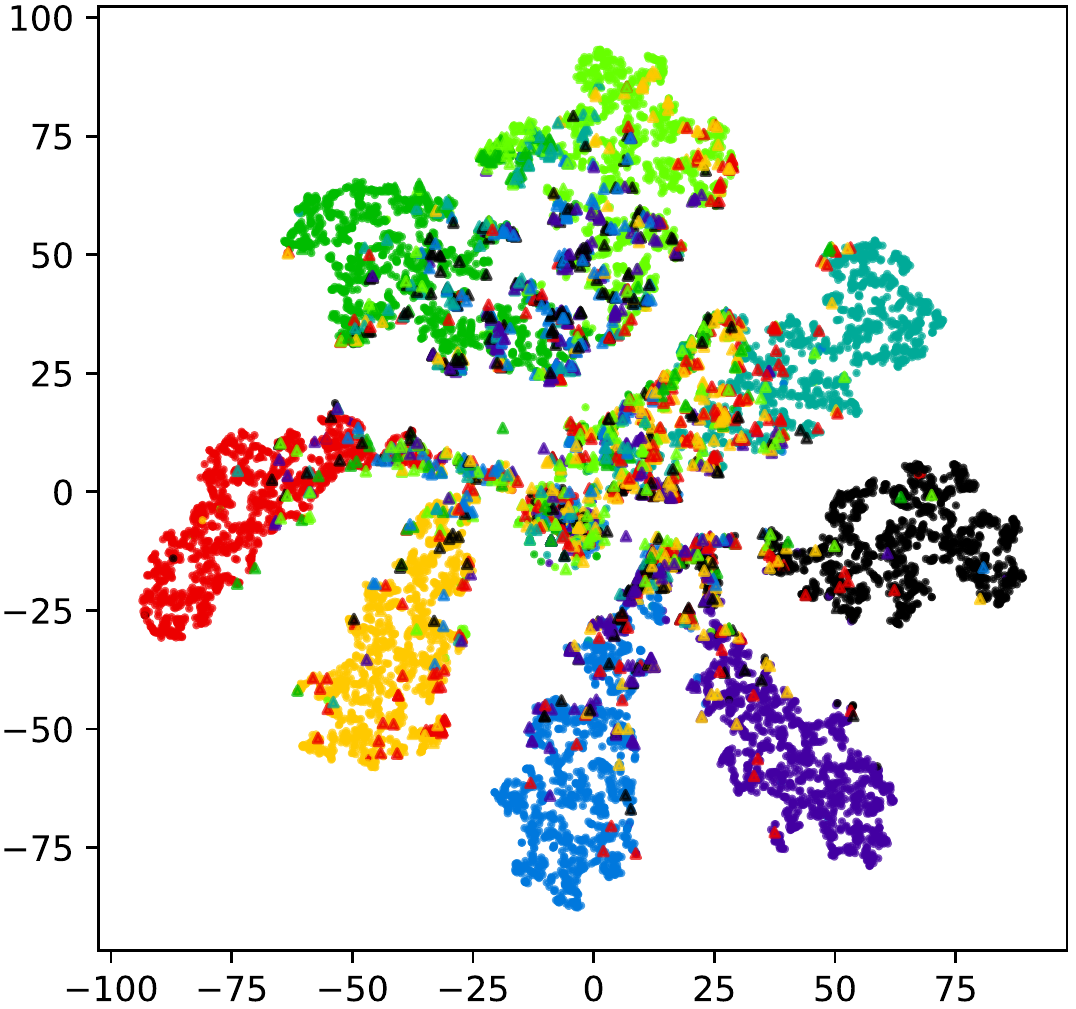}
\caption[crossmonkeytsne]{T-SNE embedding of SABLE latent space when training on 37 sessions of monkey 1 and testing on 14 sessions of monkey 2. Train trials are denoted by circles and test trials by triangles. Each colour denotes a particular movement direction.}
\label{fig:crossmonkeytsne}
\end{center}
\end{figure}

\section{Model Details}
\label{app:modeldetails}
Below are implementation details for all models used in this paper.
\begin{center}
\begin{tabular}{ |p{7cm}|p{3cm}|p{5.5cm}|  }

\hline
\multicolumn{3}{|c|}{SABLE} \\
\hline
Parameter & Value & Notes \\
\hline
Encoder &  & \\
\quad- RNN Units & 512 X 3   & Stacked Gated Recurrent Unit \\
\quad - RNN L2 Kernel Regularisation & 1000 &  \\
\quad- RNN L2 Recurrent Regularisation & 1000 & \\
\quad- Recurrent Dropout & 0.2 &  \\
\quad- $W_{enc}$ Units & 512 & Non-linear layer \\
\quad- $W_{enc}$ Dropout & 0.8 &  \\
\quad- $W_{enc}$ L2 Regularisation & 1000 &  \\
\quad- Latent space dimension & 64 &  \\
\hline
Neural Decoder &  & \\
\quad- RNN Units & 256  &  Gated Recurrent Unit \\
\quad - RNN L2 Kernel Regularisation & 0.1 &  \\
\quad- RNN L2 Recurrent Regularisation & 0.1 & \\
\quad- $W_{fac}$ Units & 128 & Non-linear layer \\
\quad- $W_{fac}$ Dropout & 0.2 &  \\
\quad- $W_{fac}$ L2 Regularisation & 10 &  \\
\hline
Behaviour Decoder &  & Batch Normalisation on all layers\\
\quad- RNN Units & 256 X 2   & Stacked Gated Recurrent Unit \\
\quad- $W_{beh}$ Units & 512 & Non-linear layer \\
\quad- $W_{beh}$ Dropout & 0.1 &  \\
\quad- $W_{beh}$ L2 Regularisation & 1.0 &  \\
\hline
Training &  & \\
Kullback–Leibler (KL) divergence weighting ($\lambda_{kl}$)   & 0.01 to 10000 & Rising exponentially 

\textbf{(between encoder and neural decoder)} \\
Reverse Gradient weighting ($\lambda_r$)   & 1.0 to 0.000000001 & Decaying exponentially \\
\hline

\end{tabular}
\end{center}

\begin{center}
\begin{tabular}{ |p{7cm}|p{3cm}|p{5.5cm}|  }

\hline
\multicolumn{3}{|c|}{SABLE-noREV} \\
\hline
Parameter & Value & Notes \\
\hline
Encoder &  & \\
\quad- RNN Units & 512 X 3   & Stacked Gated Recurrent Unit \\
\quad - RNN L2 Kernel Regularisation & 1000 &  \\
\quad- RNN L2 Recurrent Regularisation & 1000 & \\
\quad- Recurrent Dropout & 0.2 &  \\
\quad- $W_{enc}$ Units & 512 & Non-linear layer \\
\quad- $W_{enc}$ Dropout & 0.8 &  \\
\quad- $W_{enc}$ L2 Regularisation & 1000 &  \\
\quad- Latent space dimension & 64 &  \\
\hline
Neural Decoder &  & \\
\quad- RNN Units & 256  &  Gated Recurrent Unit \\
\quad - RNN L2 Kernel Regularisation & 0.1 &  \\
\quad- RNN L2 Recurrent Regularisation & 0.1 & \\
\quad- $W_{fac}$ Units & 128 & Non-linear layer \\
\quad- $W_{fac}$ Dropout & 0.2 &  \\
\quad- $W_{fac}$ L2 Regularisation & 10 &  \\
\hline
Behaviour Decoder &  & Batch Normalisation on all layers\\
\quad- RNN Units & 256 X 2   & Stacked Gated Recurrent Unit \\
\quad- $W_{beh}$ Units & 512 & Non-linear layer \\
\quad- $W_{beh}$ Dropout & 0.1 &  \\
\quad- $W_{beh}$ L2 Regularisation & 1.0 &  \\
\hline
Training &  & \\
Kullback–Leibler (KL) divergence weighting ($\lambda_{kl}$)   & 0.01 to 10000 & Rising exponentially \\
Reverse Gradient weighting ($\lambda_r$)   & N/A & \textbf{Constant positive gradient of 1} \\
\hline

\end{tabular}
\end{center}

\begin{center}
\begin{tabular}{ |p{7cm}|p{3cm}|p{5.5cm}|  }

\hline
\multicolumn{3}{|c|}{LFADS} \\
\hline
Parameter & Value & Notes \\
\hline
Encoder &  & \\
\quad- RNN Units & 512 X 3   & Stacked Gated Recurrent Unit \\
\quad - RNN L2 Kernel Regularisation & 1000 &  \\
\quad- RNN L2 Recurrent Regularisation & 1000 & \\
\quad- Recurrent Dropout & 0.2 &  \\
\quad- $W_{enc}$ Units & 512 & Non-linear layer \\
\quad- $W_{enc}$ Dropout & 0.8 &  \\
\quad- $W_{enc}$ L2 Regularisation & 1000 &  \\
\quad- Latent space dimension & 64 &  \\
\hline
Neural Decoder &  & \\
\quad- RNN Units & 256  &  Gated Recurrent Unit \\
\quad - RNN L2 Kernel Regularisation & 0.1 &  \\
\quad- RNN L2 Recurrent Regularisation & 0.1 & \\
\quad- $W_{fac}$ Units & 128 & Non-linear layer \\
\quad- $W_{fac}$ Dropout & 0.2 &  \\
\quad- $W_{fac}$ L2 Regularisation & 10 &  \\
\hline
Behaviour Decoder &  & \textbf{Trained separately to rest of model}
Batch Normalisation on all layers\\
\quad- RNN Units & 256 X 2   & Stacked Gated Recurrent Unit \\
\quad- $W_{beh}$ Units & 512 & Non-linear layer \\
\quad- $W_{beh}$ Dropout & 0.1 &  \\
\quad- $W_{beh}$ L2 Regularisation & 1.0 &  \\
\hline
Training &  & \\
Kullback–Leibler (KL) divergence weighting ($\lambda_{kl}$)   & 0.01 to 10000 & Rising exponentially \\
Reverse Gradient weighting ($\lambda_r$)   & N/A & \textbf{Constant positive gradient of 1} \\
\hline

\end{tabular}
\end{center}

\begin{center}
\begin{tabular}{ |p{7cm}|p{3cm}|p{5.5cm}|  }

\hline
\multicolumn{3}{|c|}{RAVE+} \\
\hline
Parameter & Value & Notes \\
\hline
Encoder &  & \\
\quad- RNN Units & 512 X 3   & Stacked Gated Recurrent Unit \\
\quad - RNN L2 Kernel Regularisation & 1000 &  \\
\quad- RNN L2 Recurrent Regularisation & 1000 & \\
\quad- Recurrent Dropout & 0.2 &  \\
\quad- $W_{enc}$ Units & 512 & Non-linear layer \\
\quad- $W_{enc}$ Dropout & 0.8 &  \\
\quad- $W_{enc}$ L2 Regularisation & 1000 &  \\
\quad- Latent space dimension & 64 &  \\
\hline
Neural Decoder &  & \\
\quad- RNN Units & 256  &  Gated Recurrent Unit \\
\quad - RNN L2 Kernel Regularisation & 0.1 &  \\
\quad- RNN L2 Recurrent Regularisation & 0.1 & \\
\quad- $W_{fac}$ Units & 128 & Non-linear layer \\
\quad- $W_{fac}$ Dropout & 0.2 &  \\
\quad- $W_{fac}$ L2 Regularisation & 10 &  \\
\hline
Behaviour Decoder &  & \textbf{Trained separately to rest of model}
Batch Normalisation on all layers\\
\quad- RNN Units & 256 X 2   & Stacked Gated Recurrent Unit \\
\quad- $W_{beh}$ Units & 512 & Non-linear layer \\
\quad- $W_{beh}$ Dropout & 0.1 &  \\
\quad- $W_{beh}$ L2 Regularisation & 1.0 &  \\
\hline
Domain Classifier &  & \\
\quad- Non-linear layer Units & 256 X 2   & Batch Normalisation \\
\quad- Dropout & 0.1 &  \\
\quad- L2 Regularisation & 0.001 &  \\
\hline
Training &  & \\
Kullback–Leibler (KL) divergence weighting ($\lambda_{kl}$)   & 0.01 to 10000 & Rising exponentially

\textbf{(between encoder and Domain Classifier)} \\
Reverse Gradient weighting ($\lambda_r$)   & 1.0 to 0.000000001 & Decaying exponentially \\
\hline

\end{tabular}
\end{center}
\end{document}